\documentclass{article}

\usepackage[]{units}
\usepackage{braket}
\usepackage{graphicx}
\usepackage{authblk}
\title{The ASACUSA antihydrogen and hydrogen program : results and prospects}
\author[1,2]{C.~Malbrunot}
\author[2]{C.~Amsler}
\author[2]{ S.~Arguedas Cuendi}
\author[3]{H.~Breuker}
\author[3]{P.~Dupre}
\author[2]{M.~Fleck}
\author[4]{H.~Higaki}
\author[5]{Y.~Kanai}
\author[6]{T.~Kobayashi}
\author[2]{B.~Kolbinger}
\author[6]{N.~Kuroda}
\author[7,8]{M.~Leali}
\author[2]{V.~M\"ackel}
\author[7,8]{V.~Mascagna}
\author[2]{ O.~Massiczek}
\author[6]{Y.~Matsuda}
\author[9]{Y.~Nagata}
\author[2]{M.~C.~Simon}
\author[2]{H.~Spitzer}
\author[6]{M.~Tajima}
\author[3]{S.~Ulmer}
\author[7,8]{L.~Venturelli}
\author[2]{E.~Widmann}
\author[2]{M.~Wiesinger}
\author[3]{Y.~Yamazaki}
\author[2]{J.~Zmeskal}

\affil[1]{Experimental Physics Department, CERN, Gen\`eve 23, CH-1211, Switzerland}
\affil[2]{Stefan-Meyer-Institut f\"ur Subatomare Physik, \"Osterreichische Akademie der Wissenschaften, Boltzmanngasse 3, Wien 1090, Austria}
\affil[3]{Ulmer Fundamental Symmetries Laboratory, RIKEN, Saitama 351-0198, Japan}
\affil[4]{Graduate School of Advanced Sciences of Matter, Hiroshima University, Hiroshima 739-8530, Japan}
\affil[5]{Nishina Center for Accelerator-Based Science, RIKEN, Japan}
\affil[6]{Institute of Physics, The University of Tokyo, Komaba, Meguro-ku, 153-8902 Tokyo, Japan}
\affil[7]{Dipartimento di Ingegneria dell'Informazione, Universita' di Brescia, Brescia 25133, Italy}
\affil[8]{Istituto Nazionale di Fisica Nucleare, Sez. di Pavia, I-27100 Pavia, Italy}
\affil[9]{Department of Physics, Tokyo University of Science, Shinjuku, Tokyo 162-8601, Japan}
\date{}                     
\begin{document}
\maketitle
\begin{abstract}
The goal of the ASACUSA-CUSP collaboration at the Antiproton Decelerator of CERN is to measure the ground-state hyperfine splitting of antihydrogen using an atomic spectroscopy beamline. A milestone was achieved in 2012 through the detection of 80 antihydrogen atoms 2.7 meters away from their production region. This was the first observation of ``cold'' antihydrogen in a magnetic field free region. In parallel to the progress on the antihydrogen production, the spectroscopy beamline was tested with a source of hydrogen. This led to a measurement at a relative precision of  $2.7\times 10^{-9}$ which constitues the most precise measurement of the hydrogen hyperfine splitting in a beam.  Further measurements with an upgraded hydrogen apparatus are motivated by CPT and Lorentz violation tests in the framework of the Standard Model Extension. \\
Unlike for hydrogen, the antihydrogen experiment is complicated by the difficulty of synthesizing enough cold antiatoms in ground-state. The first antihydrogen quantum states scan at the entrance of the spectroscopy apparatus was realized in 2016 and is presented here. The prospects for a ppm measurement are also discussed.
\end{abstract}
 
\section{Introduction}\label{s:intro}

Since the first detection of relativistic antihydrogen atoms more than 20 years ago at LEAR (CERN) \cite{Baur} and later at Fermilab \cite{Blanford}, the field of antihydrogen research  rapidly took momentum with the start of the Antiproton Decelerator (AD) at CERN in 2000. The first detection of low energy antihydrogen was reported in 2002 \cite{Amoretti, Gabrielse} followed by the first magnetic trapping of antihydrogen in 2010 \cite{trapping} which enabled the first measurements on trapped antihydrogen atoms in the following years. Given the magnetic moment of antihydrogen in ground-state, trapping requires typical temperatures smaller than $\sim$\unit[0.5]{K} which is challenging since the adopted formation mechanism requires the interaction of trapped antiproton and positron clouds. The ASACUSA-CUSP collaboration proposed in 2005 a measurement of the ground-state hyperfine splitting of antihydrogen using a beam method allowing a relaxed constraint, of about 2 orders of magnitude, on the temperature of the antihydrogen atoms available for measurements \cite{ASACUSAproposal, Mohri}.\\ 

The race towards producing large amount of cold antihydrogen atoms is motivated by the appealing prospects for CPT symmetry tests. The measured atomic transitions on hydrogen, one of the best studied atomic systems, constitute a precise comparison ground for antihydrogen. A direct consequence of the CPT theorem \cite{Bell, Luders, Luders2, Pauli} is that antihydrogen and hydrogen should have the exact same spectrum. Measuring atomic transitions in antihydrogen with high precision therefore promises one of the most stringent tests of CPT symmetry. The motivation behind testing such a cornerstone of quantum field theory is manifold. The baryon asymmetry in the universe reflected by the notable absence of primordial antimatter remains to-date unexplained. Additionally, quantum field theory, although extremely successful, yet fails to include the gravitational force. \\

Tests of CPT symmetry have been and are being performed in several different physical systems. In the leptonic sector, the symmetry has been tested  by comparison of the electron and positron as well has the charged muons g-factors \cite{CPTlepton, CPTmuons}, in baryons with recently the most precise CPT test on baryonic systems performed on antiprotons \cite{CPTbaryon}, in mesons with the famous neutral kaons mass comparison \cite{CPTmeson} and even in nuclei with the recent measurement in ALICE of the charge-to-mass ratio of anti-helion and anti-deuteron \cite{CPTnuclei}.  The so far only accessible atomic system purely consisting of antimatter is antihydrogen, for which  
the first optical transitions between the 1S and 2S states were observed paving a path to precise measurements \cite{ALPHA_1S2S}. The ground-state hyperfine splitting has been very recently measured with a relative precision of  $4\times10^{-4}$ \cite{GSHF_ALPHA}.  Relative precisions is commonly used to compare the sensitivity of experiments. In the context of CPT symmetry tests, the Standard Model Extension (SME) \cite{SME} sets a framework in which experiments searching for CPT and Lorentz symmetry violations can be compared to each other. In this context, CPT violation arises from the inclusion of all possible effective operators for Lorentz violation to the standard model Lagrangian.  A Lorentz-violating term in the Lagrangian of the SME is constructed from a tensor coefficient contracted with a conventional tensor operator. The coefficients act as Lorentz-violating background fields. For most low-energy systems the absolute energy scale probed by the experiments defines their sensitivity to those fields. 
For antihydrogen \cite{Bluhm, Vargas}, the 1S-2S transition and the ground-state hyperfine splitting, to cite only the most precisely measured transitions in hydrogen, are sensitive to different coefficients of the SME fields and can therefore provide different constraints to potential new physics.\\

The antihydrogen trap experiments' early focus was dedicated to the 1S-2S transition \cite{ALPHAproposal, ATRAPproposal} which was measured for hydrogen in a trap with a relative precision of $2\times10^{-12}$ \cite{Cesar}. An almost 3 orders of magnitude more precise value was achieved in a beam in 2011 \cite{Parthey}. The ASACUSA-CUSP collaboration intends to measure the ground-state hyperfine splitting of antihydrogen in a beam. The same transition in hydrogen was measured in a beam (before the new ASACUSA-CUSP measurement, see \ref{s:hydrogen}\ref{s:results}) with a precision of $5\times10^{-8}$ \cite{Prodell, Kusch}, a 4 orders of magnitude more precise value was achieved in a maser in the 1970s \cite{Hellwig, Karshenboim, Essen, Ramsey1, Ramsey2}.\\

The advantage of the beam setup for the measurement of the hyperfine splitting of antihydrogen is that, on top of having a larger temperature acceptance than traps, it allows a measurement in a near magnetic field-free region. Indeed the hyperfine splitting being a magnetic phenomenon, the measurable transitions between the different hyperfine states are very sensitive to magnetic field gradients. Thus, the technique in principle allows for better resolution, accuracy and precision on the hyperfine splitting than a measurement in a trap. 
The challenges however lie in forming a polarized beam of ground-state antihydrogen atoms.\\
In the first place a focussed beam of antihydrogen should be produced to compensate for the loss of solid angle due to the distance between the production and the detection points. In the current design of the ASACUSA-CUSP antihydrogen apparatus \cite{Sauerzopf}, the antihydrogen detector sees a solid angle of \unit[$5\times10^{-4}$]{sr}. Even the most efficient production of antihydrogen, which nowadays converts $\sim30\%$ of the trapped $\sim$10$^5$ antiprotons, would only achieve a couple of antihydrogen at the detector per trial without a beam formation. Secondly, the beam should be polarized. In a trap the polarization is automatic as the untrappable states (called high-field seekers) annihilate on the surrounding electrodes right after their production and only the trappable states (two low-field seeker states with  total angular
momentum quantum number $F$ and magnetic quantum number $M_{F}$ equal to ($F$,$M_F$)=(1,0) and ($F$,$M_F$)=(1,1) (the latter being ($F$,$M_F$)=(1,-1) for antihydrogen) remain in the trap. A third challenge in a beam method relates to the quantum states of the antihydrogen atoms. Three-body recombination, which is the dominating production mechanism in the ASACUSA-CUSP experiment \cite{Enomoto}, mostly produces antihydrogen in highly excited states \cite{Robicheaux}. They spontaneously decay to ground-state within a ns (for low n) to ms (for e.g. circular states of n$\sim$30) time scale. 
The temperature acceptance of the ASACUSA-CUSP apparatus is limited to roughly \unit[50]{K}, which translates into velocities of $\sim$\unit[1000]{m/s}.
Such velocities in a beam do not allow enough time for decays from high-lying n-states (n>25 \cite{Lundmark}) to happen before the atoms have reached the spectroscopy apparatus. In contrast, atoms can remain in a trap for tens of minutes. Therefore, in a beam, the production of ground-state, or nearly ground-state, atoms needs to be enhanced or stimulated deexcitation mechanisms need to be implemented.\\

We have underlined here, the experimental challenges specific to the ground-state hyperfine splitting measurement of antihydrogen in a beam and we will discuss in section \ref{s:hbar} how they are being addressed by the ASACUSA-CUSP collaboration. In the next section we will describe the results yielded by a test apparatus used on hydrogen and motivate further measurements in hydrogen in the context of the SME.\\

\section{Hyperfine splitting measurement}\label{s:hydrogen}
\subsection {Theoretical background}\label{s:theory}
The hyperfine splitting in (anti)hydrogen arises from the interaction of the magnetic moments of the electron (positron) and proton (antiproton). 
To first order it is proportional to the product of those. The magnetic moment of the antiproton is known since a few years to the ppm level\cite{mag_moment_Ulmer, mag_moment_Jack}. At the 10$^{-5}$ level \cite{Friar}, corrections to the first order calculation of the hyperfine splitting introduce contributions due to the magnetic and electric form factors of the antiproton which are to date unknown. Therefore, given the current knowledge on the magnetic moment of the antiproton, the ground-state hyperfine splitting of antihydrogen is sensitive at the \unit[30]{ppm} level to the structure of the antiproton. This strongly motivates, in addition to the prospect for more sensitive CPT tests, further experiments beyond the currently achieved relative precision of $4\times10^{-4}$ in a trap.

\begin{figure}[!h]
\centering\includegraphics[width=\textwidth]{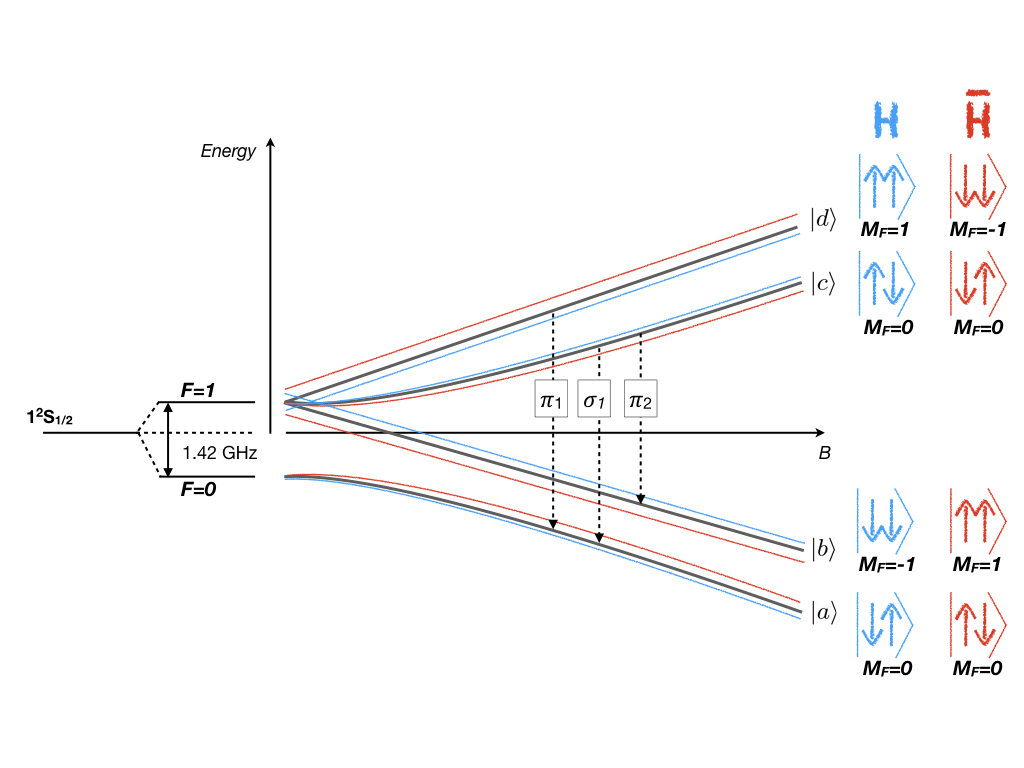}
\caption{Illustrative figure of the Breit-Rabi diagram and the potential effect of the SME's CPT and Lorentz violating fields on the hyperfine splitting of hydrogen and antihydrogen.  The  sign and magnitude of the shift affecting each states within hydrogen (or antihydrogen) are dependent on the magnitude of the  $g$ and $H$ coefficients in equation \ref{eq:frequency-shift} as well as  the $c$ and $a$ coefficients mentioned in the text \cite{Vargas, spinweight}. Here we have assumed that the effect on hydrogen and antihydrogen are opposite and of different, non-zero, amplitude which at least implies that $g\neq H$.}
\label{fig_splitting}
\end{figure}

In presence of an external magnetic field the hyperfine levels are further split due to the additional interaction term proportional to $\mu B$  (where $\mu$ is the hydrogen magnetic moment and $B$ the magnetic field) in the Hamiltonian. Fig.\ref{fig_splitting} illustrates how the $F=1$ triplet state's degeneracy is lifted  in a magnetic field and the dependence of the states on the magnetic field's amplitude. In this figure the $\ket{d}$ and  $\ket{c}$ states refer to the low-field seeking and  $\ket{b}$ and  $\ket{a}$ to the high-field seeking states mentioned above. Three transitions between low field seeking and high field seeking states are then possible. The $\pi_1$ and $\pi_2$  (for which $\Delta M_{\textrm{F}}=\pm1$) as well as the $\sigma_1$ ($\Delta M_{\textrm{F}}=0$) transitions allow for the determination of the zero-field transition which can be compared to theoretical calculations (to the $10^{-5}$ level as noted above) or to other experimental measurements. 
The zero field value can be extracted through extrapolations by measuring the $\pi_1$ or $\sigma_1$ transitions at different external magnetic field and using the Breit-Rabi formula that relates the energy of the states to the external magnetic field value \cite{Breit}. 
Alternatively one can measure two transitions  at the same magnetic field to directly compute the zero-field value.
In the presence of Lorentz and CPT violating effects, those two methods could lead to different results.  One should note that the $\sigma$ transition at zero-field is not sensitive to SME fields because the shift of the hyperfine transitions is proportional to $\Delta M_{\textrm{F}}$ (see equation \ref{eq:frequency-shift}). In general, computations which cancel the effect of the linear Zeeman shift also cancel the effect of Lorentz and CPT violating terms. This is the case of the difference between the $\pi_1$ and $\pi_2$ transitions which does not lead to any constraints on SME coefficients. Fig.\ref{fig_splitting} illustrates the potential effect of SME fields on hydrogen and antihydrogen hyperfine structures. The magnitude and the sign of the effect for hydrogen and antihydrogen is unknown and depends on the relative strength of CPT-odd and CPT-even terms. In order to determine those, measurements on both matter and antimatter are necessary and should be realized at the same time and at the same location.\\
In hydrogen additional constraints on SME coefficients can be obtained by measuring sidereal variations of the hyperfine splitting which could be caused by the change of the magnetic field orientation (due to the rotation of the Earth) with respect to the background fields responsible for Lorentz violation. Similar constraints could be obtained in antihydrogen if the rate of antiatoms would permit such fast measurements. 
SME coefficients measured in the laboratory frame can be expressed in terms of coefficients in the Sun-centered frame (in which the Lorentz breaking background fields are assumed to be constant) as \cite{Vargas} :

{\small
\begin{equation}
	\mathcal{K}_{w_{k10}}^{Lab} = \mathcal{K}_{w_{k10}}^{Sun} \cos(\theta) - \sqrt{2}\ \Re e(\mathcal{K}_{w_{k11}}^{Sun}) \sin(\theta) \cos(\omega_\oplus T_\oplus) + \sqrt{2}\ \Im m(\mathcal{K}_{w_{k11}}^{Sun}) \sin(\theta) \sin(\omega_\oplus T_\oplus)  ,\label{eq:siderialvariations}
\end{equation}
} 

where $\omega_\oplus=\frac{2\pi}{23h56min}$ is the earth's rotation frequency, $T_\oplus$ is the sidereal time and $\theta$ is the angle between the applied magnetic field and the earth's rotational axis. 
The symbol $\mathcal{K}_{w_{k10}}^{Lab}$  denotes either of the following coefficients $g^{0B}$, $H^{0B}$, $g^{1B}$, $H^{1B}$ which are non-relativistic spherical coefficients for Lorentz violation expressed in the laboratory frame. For clarity, we have omitted here the superscript $NR$ on every coefficients but it should be noted that all SME coefficients mentioned in this manuscript are non-relativistic. The 0 an 1 superscripts refer to the spin weight \cite{spinweight} and the $w$ subscript stands for electron or proton. The index $k$ represents the mass dimension of the coefficient. Here, we restrict to $k\leq4$ given that the coefficients are suppressed by $(\alpha m)^{2k}$ . The $g$ coefficients control CPT-odd effects, while the $H$ ones control CPT-even effects. Each specific coefficient governs a physically distinct violation of Lorentz symmetry. Two additional coefficients, $c$ and $a$ are involved in the shift of the hyperfine states. However they shift all hyperfine sub-levels of a principal state by the same amount and are therefore not discussed here, as they are not accessible by direct observation of the hyperfine splitting. \\
In the laboratory frame the shift of an hyperfine transition is given by :
{\small
\begin{equation}
	2\pi\delta\nu(\Delta M_F) =  \frac{\Delta M_F}{2\sqrt{3\pi}} \sum_{q=0}^{2} (\alpha m_r)^{2q} \left( 1+ 4 \delta_{q2}  \right) \sum_{w} \left[- g_{w_{(2q)10}}^{0B} + H_{w_{(2q)10}}^{0B} - 2 g_{w_{(2q)10}}^{1B} + 2 H_{w_{(2q)10}}^{1B} \right].
\label{eq:frequency-shift}
\end{equation}}
Here, $k=2q$ and therefore $q$ spans from 0 to 2. Since $w$ = e,p and there are four types of non-relativistic coefficients, we count $3\times2\times4 = 24$ independent coefficients contributing to the frequency shift of the $\pi$ transition in the laboratory frame.
Each laboratory-frame coefficient being associated with three independent coefficients in the Sun-centered frame (equation \ref{eq:siderialvariations}) there are a total of $24\times3 = 72$ independent SME coefficients in the Sun-centered frame.

The measurement of the hyperfine transition using a maser reached an absolute precision of mHz \cite{maser1, maser2, maser3}. Siderial measurements at the same precision led to the constraint of 48 of the coefficients mentioned above (the ones embedded in $\mathcal{K}_{w_{k11}}^{Sun}$) setting bounds at the level of \unit[$2\times 10^{-27}$]{GeV} \cite{maser3}. The beam technique described in the next section can reach a precision of the order of a Hz and  therefore cannot provide better constraints than the maser measurement on sidereal coefficients. However the remaining 24 unconstrained coefficients ($\mathcal{K}_{w_{k10}}^{Sun}$) can be probed by changing the field orientation in the laboratory. \\

We have so far ignored contributions due to the rotation of the Earth around the Sun. If one considers this additional boosted and rotated frame, additional oscillations in the signal could be observed. Performing measurements at different times of the year would enable constraining additional coefficients which are to-date unprobed (but are however suppressed with respect to the ones listed above \cite{Vargas}).

\subsection{Latest results on Hydrogen}\label{s:results}
The measurement of the $\sigma_1$ transition to an unprecedented precision in a beam was recently reported in \cite{Diermaier}. The spectroscopy apparatus (a resonant cavity in which  the hyperfine transition is driven and a superconducting sextupole magnet to select the spin state)  designed for the ASACUSA-CUSP antihydrogen experiment (see \S\ref{s:status}) was used for this measurement. 

The $\sigma_1$ transition, having a quadratic dependency on the  external static magnetic field, is less sensitive to the inhomogeneity of the field and could therefore be measured using a single pair of Helmholtz coils providing a field of a few Gauss in a direction perpendicular to the beam and parallel to the RF field, the latter being necessary to drive the  $\sigma$ transition, and an homogeneity of  $\frac{\sigma_B}{B}\sim1\%$ at the cavity. An uncertainty of a few Hz was reached (\unit[2.7]{ppb} relative precision) and no significant signs for systematic errors have been encountered at this level of precision. However, as mentioned in \S\ref{s:theory},  $\sigma_1$ is insensitive to Lorentz and CPT violating SME fields and a new measurement campaign was started to measure with a similar precision the $\pi_1$ transition. At \unit[10]{Gauss} the $\sigma_1$ and $\pi_1$ transitions differ by $\sim$\unit[14]{MHz}. The cavity used is resonant at \unit[1420]{MHz} with a bandwidth of \unit[12]{MHz} allowing the measurement of both transitions.\\

\begin{figure}[!h]
\centering\includegraphics[width=1.00\textwidth]{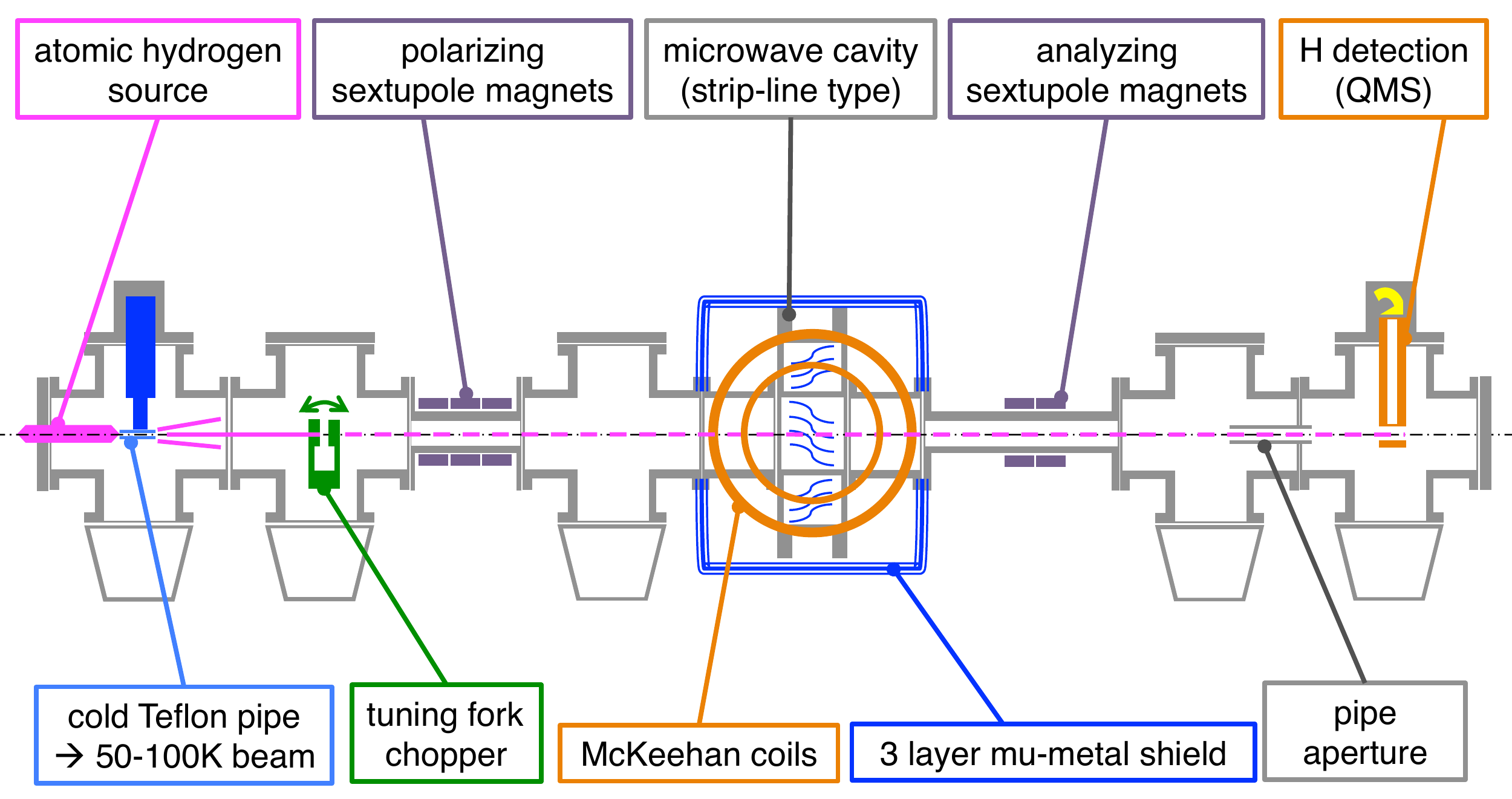}
\caption{Sketch of the hydrogen experiment setup to measure the $\pi_1$ transition. }
\label{fig_happaratus}
\end{figure}

We report here the first observation of the $\pi_1$ transition in a beam. This measurement required an additional set of correction coils and a new magnetic shielding design which combined provided a field homogeneity better than $\frac{\sigma_B}{B}\sim0.1\%$. The cavity is rotated by 45$^{\circ}$ around the hydrogen beam axis to allow both the  $\sigma$ and $\pi$ transitions to be driven (for the latter one, the external magnetic field needs to be perpendicular to the RF field).
Fig.\ref{fig_happaratus} shows the experimental setup used for this measurement. The cavity and the two pairs of Helmholtz coils (in a so-called McKeehan configuration \cite{McKeehan}) are enclosed in the magnetic shielding. The cavity is identical to the one previously used in the $\sigma_1$ determination. Its ``strip-line" design leads to a lineshape with a double-dip structure (see \cite{Diermaier} for more details on the lineshape's structure). The superconducting sextupole, being used in the antihydrogen experiment detailed in \S\ref{s:hbar}, was replaced by sets of permanent magnets with a smaller diameter but a similar integrated gradient in order to allow both hydrogen and antihydrogen experiments to be operated independently. 

\begin{figure}[!h]

\centering\includegraphics[width=\textwidth]{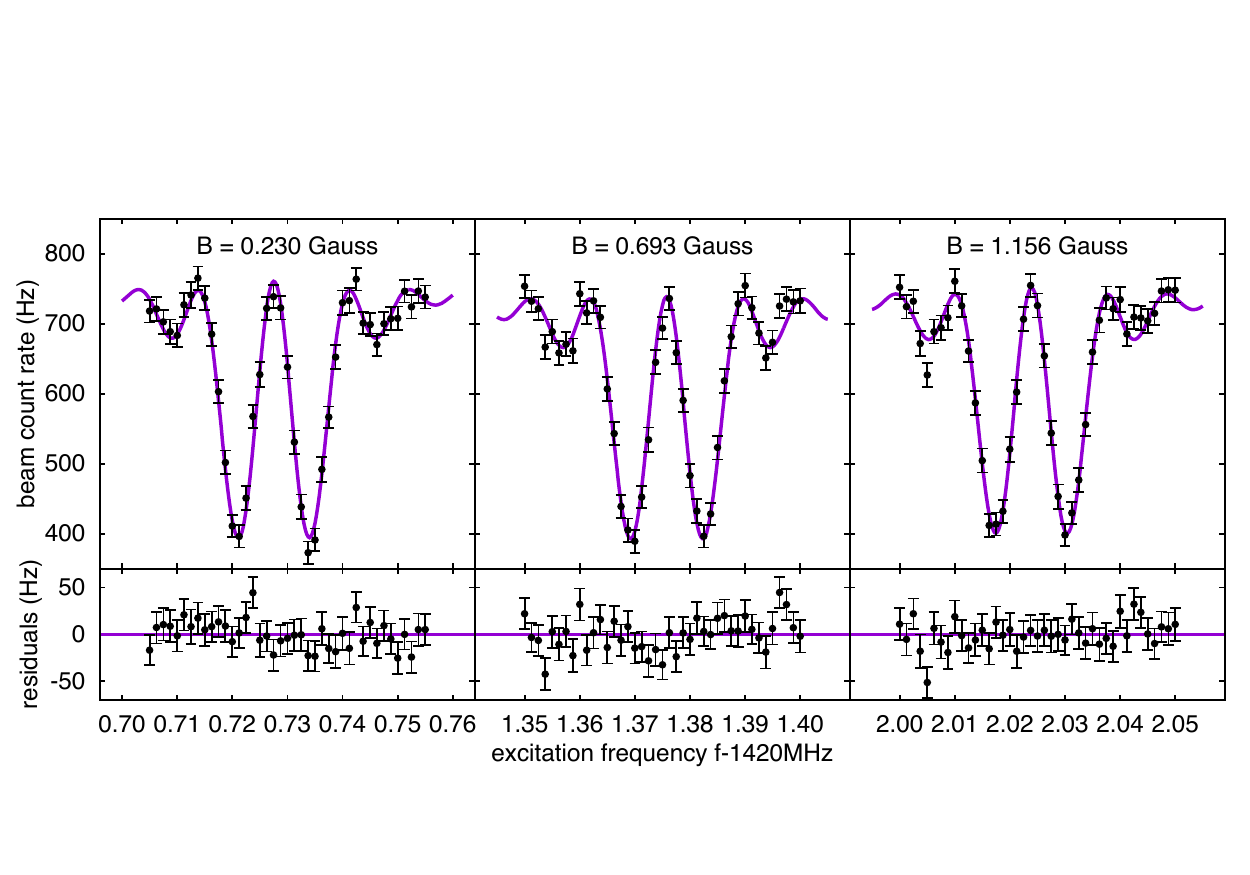}
\caption{$\pi_1$ transitions observed at different external magnetic field amplitudes. 41 data points are taken over a scan range of \unit[50]{kHz}. The lineshape is fitted to extract the central frequency with a $\sim$\unit[60]{Hz} precision. Residuals are shown in the bottom plots.}
\label{fig_pi1}
\end{figure}

Fig.\ref{fig_pi1} shows the $\pi_1$ transition measured at 3 different fields.
Since the homogeneity of the external field allows the resolution of the double-dip structure, the precision of this method is dominated by the interaction time of the atoms with the microwave field. A velocity of \unit[1000]{ms$^{-1}$} leads to a linewidth of \unit[12]{kHz}. With an acquisition time of \unit[40]{min} we achieved a precision on the central frequency, which splits the observed linewidth by a factor of ~200. A few Hz precision can be reached by measuring both $\pi_1$ and $\sigma_1$ transitions within a one week-long measurement campaign \cite{Sergio}.

\subsection{Outlook}

This result shows that new constraints on  (or determination of!) SME coefficients can be made in the near future. Measurements of the $\pi_1$ transition at opposite B-field directions using the $\sigma_1$ transition  (which is insensitive to SME fields) as a reference will be performed to access the 24 unconstrained coefficients mentioned in \ref{s:hydrogen}\ref{s:theory}.
Comparison of the zero-field frequency obtained using the extrapolation method on the $\pi_1$ transition and the combination of the $\pi_1$ and $\sigma_1$ transitions will provide additional constraints.
In a second step, further measurements at different times of the year will enable assessing other un-constrained coefficients. It is also worth noting that the hydrogen experiment is located at the same earth's latitude and longitude coordinates than our analog antihydrogen experiment.\\
 
\section{Antihydrogen measurement}\label{s:hbar}

As already mentioned, antihydrogen spectroscopy will further constrain the SME landscape in particular the CPT-odd terms. 
The recent measurement of the hyperfine splitting in hydrogen enabled an estimation of the number of antihydrogen needed to reach a ppm relative precision using the same technique. Few assumptions on the antihydrogen beam properties were made which included an estimation of atoms in excited states in the cavity. 
An important difference between the hydrogen and antihydrogen experiment is indeed that antihydrogen atoms are formed in highly excited states. A determination of the quantum state distribution of antihydrogen atoms at the entrance of the spectroscopy apparatus was therefore necessary. It was performed using an external field-ionizer (described below) upstream of the cavity to reject atoms in a high principal quantum number state. The result of the measurement done in the ASACUSA-CUSP antihydrogen apparatus is presented and discussed in \S\ref{s:qs}.

\subsection{Status}\label{s:status}
\begin{figure}[!h]
\centering\includegraphics[clip, trim=0.cm 6cm 0cm 6cm, width=1.00\textwidth]{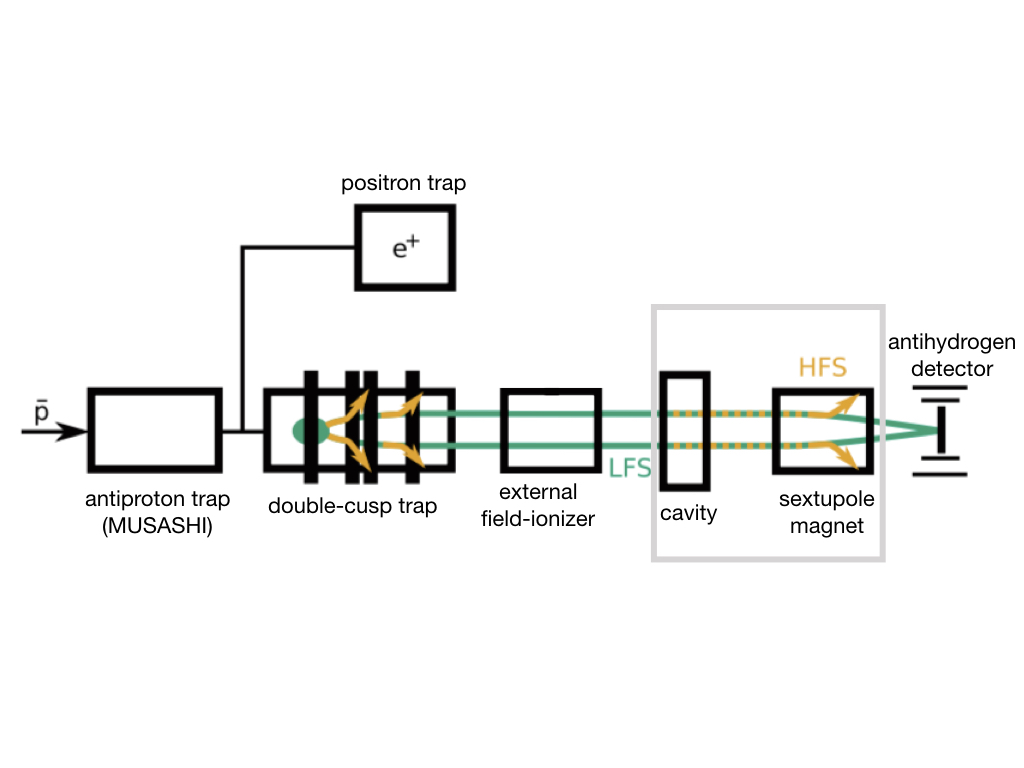}
\caption{Sketch of the ASACUSA-CUSP antihydrogen setup. The spectroscopy apparatus used in the hydrogen $\sigma_1$ measurement reported in \cite{Diermaier} is highlighted by the grey box. For the quantum state measurements the highlighted components were removed, and the antihydrogen detector was placed directly downstream of the external field-ionizer.}

\label{apparatus_hbar}
\end{figure}

The first measurements of antihydrogen atoms in a field-free environment, \unit[2.7]{m} away from the production region was reported by the ASACUSA-CUSP collaboration in 2014 \cite{Kuroda}.
Since then, efforts were concentrated on increasing the flux and characterizing  beam properties in view of the spectroscopic measurements. For this latter purpose a field-ionizer capable of ionizing states down to principle quantum number n$\sim$14 \cite{Sauerzopf} was added directly upstream of the cavity. Together with the field-ionizer internal to the CUSP and closer to the production region it provided a diagnostic on the states of the antihydrogen produced inside and exiting the CUSP. During the 2016 data-taking period, the antihydrogen detector seen in Fig.\ref{apparatus_hbar} was placed directly downstream of the field-ionizer. The detector consists of a central BGO calorimeter readout by multi-anode PMTs \cite{NagataBGO} (therefore additionally providing position resolution) and a 2 layers hodoscope made of 32 scintillator bars each, read out on both side by silicon photomultipliers \cite{Clemens}. 
The combination of the vertexing capability of the detector and the measurement of the energy deposit at the annihilation point provides a strong discriminating power between cosmic (the main background component) and antihydrogen signals.
Fig \ref{apparatus_hbar} shows the ASACUSA-CUSP antihydrogen apparatus. Synthesis of antihydrogen is done in a double-cusp trap inside which multi-ring electrodes provide the electrostatic field necessary to trap the charged particles axially. The double-cusp trap provides a strong magnetic field gradient \cite{Nagata_cusp} which should enhance the polarizing  and focussing effect on the exiting antihydrogen beam. 

\subsection{Measurement of the quantum state distribution}\label{s:qs}
\begin{figure}[!h]
\centering\includegraphics[width=4in]{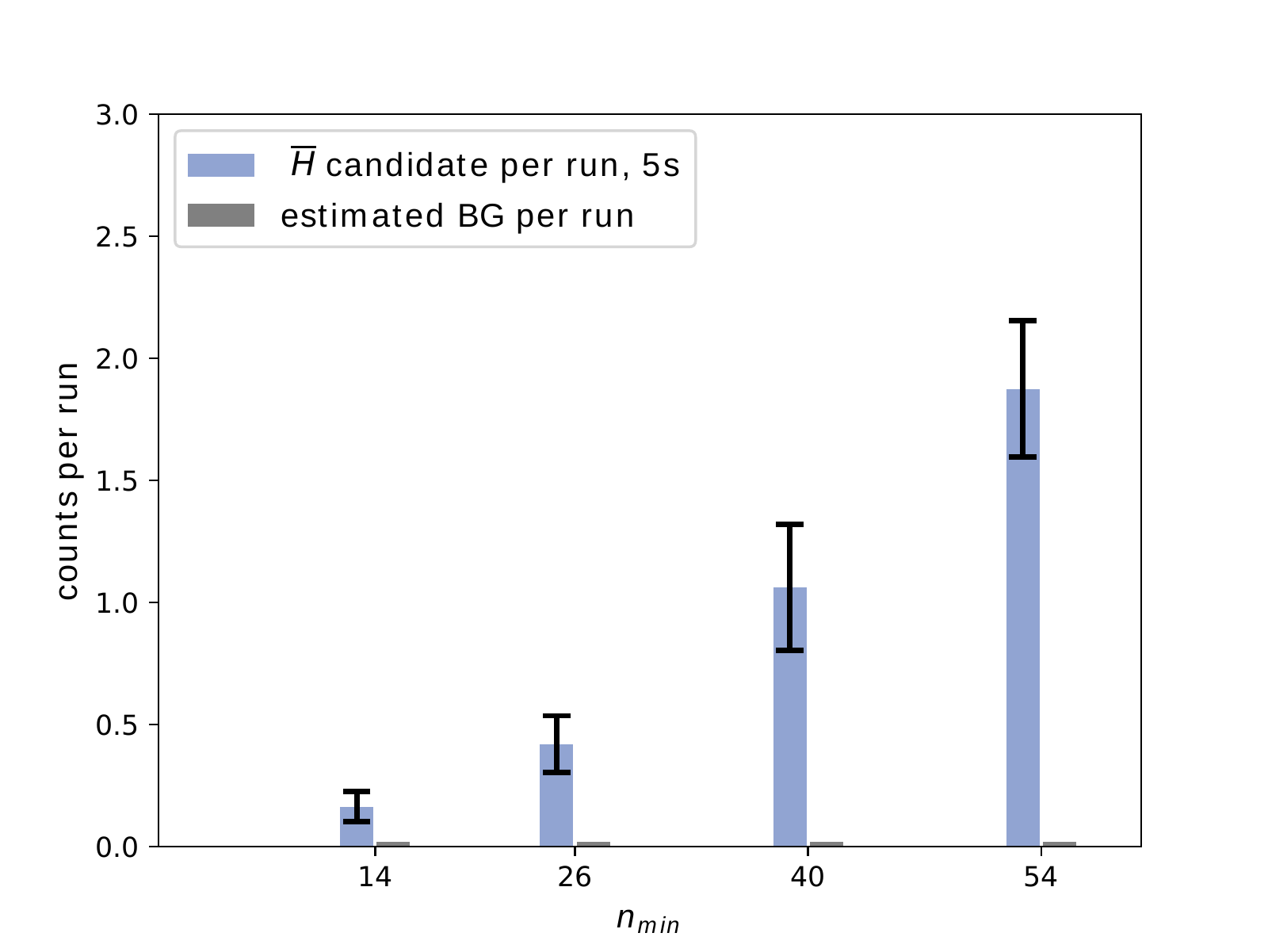}
\caption{Measurement of the quantum state distribution of antihydrogen atoms at the entrance of the spectroscopy apparatus. The indicated quantum number n$_\textrm{min}$ is the minimum principal quantum number ionized. The counts on the detector at that particular n$_\textrm{min}$ therefore include all atoms having a principal quantum number n$<$n$_\textrm{min}$. The counts are averaged over the first \unit[5]{s} after the start of the mixing of antiprotons and positrons in the double-cusp trap. Poisson errors on the antihydrogen candidates are indicated. The errors on the background events are estimated by averaging the algorithm's outcome to 500 randomly chosen training samples. They are too small to be seen on the graph.}
\label{fig_qdist}
\end{figure}

We present here the first quantum state distribution measurement of antihydrogen atoms in a low magnetic field region. The analysis of the data was done using gradient boosted decision trees, a machine learning algorithm, which was trained on antiprotons extracted from the MUSASHI trap towards the detector. The annihilation pattern of those antiprotons is from the detector's point of view identical to those of antihydrogen. About 4000 annihilations of antiprotons were recorded. 2/3 of those were used to train the algorithm and the other third was used to test the algorithm and extract the signal efficiency. Additionally about 30000 cosmic events were recorded and 2/3 used to train the algorithm on background recognition. The efficiency of the algorithm in detecting antiproton is close to \unit[80]{\%} while the rate of mis-identified antiproton events in the cosmic sample is less than \unit[0.25]{\%} (The details of the analysis procedure will be published elsewhere \cite{Bernadette_tobepublished}). The angle of the tracks (tracks are defined as originating from the BGO detector) recorded in the detector with respect to the horizontal axis and the energy deposited in the BGO detector turned out to be the most important features for discriminating annihilation events from cosmic background. \\
The algorithm was then used to identify antihydrogen events during mixing runs with four different settings of the external field-ionizer. The results of the analysis, normalized to a run, is shown in Figure \ref{fig_qdist}. The cosmic trigger rate in the detector is of the order of \unit[1.6]{Hz} so that in a mixing run of \unit[5]{s}, approximately 8 cosmic events are expected leading to 0.02 fake antihydrogen events. The smallest number of antihydrogen events were recorded for the highest field-ionizer configuration ($\vec{E}\sim\unit[10]{kV/cm}$) and averaged to about \unit[0.16]{events} per run. Given the cosmic rejection, the background rate is more than 8 times smaller. The antihydrogen atoms which are not ionized by the strongest field of the field-ionizer have a principal quantum number smaller than 14. From this state, the longest decay channel to ground state is of the order of \unit[100]{micro-seconds}. Assuming velocities of \unit[1000]{m/s} (which is roughly the acceptance limit of the apparatus), antihydrogen atoms in those states would be in the ground state before they reach the cavity, apart from atoms decaying to the metastable state 2S. Within 43 mixing runs, 7  antihydrogen events with principal quantum number n$<14$ were recorded with a significance of \unit[4.5]{$\sigma$}.   

\subsection{Prospects}
The first measurement of the antihydrogen quantum state distribution down to low quantum states confirms the higher proportion of high Rydberg states at the exit of the production trap. At this stage, the observed rate of low lying states is too small to reach the necessary number of antiatoms for a ppm measurement. Efforts are now concentrated on stimulating the deexcitation close to the production point as well as enhancing the production of ground-state antihydrogen. For that purpose different mixing schemes are being developed and   positron's density and temperature are being optimized.\\
The Extra Low Energy Antiproton Ring ELENA \cite{ELENA} which is being commissioned at the AD, will provide, starting from 2021 for the majority of antihydrogen experiments, a lower beam energy and a higher beam availability which will be beneficial to  the ASACUSA-CUSP experiment on two fronts : a round-the-clock antiproton availability  which will avoid the daily time consuming beam-tuning through the ASACUSA RFQD and a separate beamline for the second ASACUSA activity, sparing the bi-annual disassembly and assembly of the entire apparatus and therefore allowing for necessary developments throughout the year.

\section{Conclusions and Outlook}
We have presented the latest results of the hydrogen and antihydrogen experiments of the ASACUSA-CUSP collaboration. The recent hydrogen result followed by the first observation of the $\pi_1$ transition opens the way to further measurements which will provide additional constraints on SME coefficients. In the antihydrogen experiment, first atoms with a low principal quantum number were detected at the entrance of the spectrometer. This result calls for a stimulated deexcitation in order to reach a rate of $\sim$ 10 ground-state counts per run which would be compatible with a ppm measurement.

\enlargethispage{20pt}

\section*{Author's contributions}
C.M, S.A.C, B.K, O.M, M.C.S, M.W., E.W and J.Z are involved in the hydrogen experiment described in \S\ref{s:hydrogen}. C.M, C.A., H.B., P.D., M.F. H.H, Y.K., T.K., B.K, N.K., M.L., V.M\"a, V.M., O.M, Y.M., Y.N, M.C.S., M.T., S.U., L.V, E. W, Y.Y, and J.Z. are involved in the antihydrogen experiment  described in \S\ref{s:hbar}.  B.K analyzed the data presented in \S\ref{s:qs} (an independent analysis, not discussed here, which is not relying on machine learning was performed by Y.N.). C.M wrote the manuscript which was discussed among all authors.

\section*{Acknowledgments}
We express our gratitude towards the AD group of CERN. We would like to thank A. Vargas for fruitful discussions on the Standard Model Extension.\\
This work has been supported by the European Research Council under European Union's Seventh Framework Programme (FP7/2007-2013)/ERC Grant agreement (291242), the Austrian Ministry of Science and Research, the Austrian Science Fund (FWF): W1252-N27, the Grant-in-Aid for Specially Promoted Research (24000008) of MEXT and the RIKEN Pioneering Project.


\end{document}